\documentclass[%
 aip,
 amsmath,amssymb,
 reprint,%
]{revtex4-1}
\usepackage{comment}
\usepackage{graphicx}
\usepackage{dcolumn}
\usepackage{bm}

\usepackage[utf8]{inputenc}
\usepackage[T1]{fontenc}
\usepackage{mathptmx}
\usepackage{etoolbox}
\usepackage{xcolor}
\usepackage{soul}

\makeatletter
\def\@email#1#2{%
 \endgroup
 \patchcmd{\titleblock@produce}
  {\frontmatter@RRAPformat}
  {\frontmatter@RRAPformat{\produce@RRAP{*#1\href{mailto:#2}{#2}}}\frontmatter@RRAPformat}
  {}{}
}%
\makeatother
\begin{document}

\preprint{AIP/123-QED}

\title{Physics-based Full-band GaN High-Electron-Mobility Transistor Simulation Suggests Upper Bound of LO Phonon Lifetime}

\author{Ankan Ghosh Dastider}
\affiliation{%
Department of Electrical and Computer Engineering, University of Illinois Urbana-Champaign, Urbana, IL, USA
}%
\affiliation{%
Holonyak Micro and Nanotechnology Laboratory, University of Illinois Urbana-Champaign, Urbana, IL, USA
}%

\author{Matt Grupen}
\affiliation{ 
Air Force Research Laboratory Sensors Directorate, 2241 Avionics Cir., Wright-Patterson AFB, OH, USA
}%

\author{Ashwin Tunga}
\affiliation{%
Department of Electrical and Computer Engineering, University of Illinois Urbana-Champaign, Urbana, IL, USA
}%
\affiliation{%
Holonyak Micro and Nanotechnology Laboratory, University of Illinois Urbana-Champaign, Urbana, IL, USA
}%

\author{Shaloo Rakheja\textsuperscript{*,}}
  \email{rakheja@illinois.edu}
\affiliation{%
Department of Electrical and Computer Engineering, University of Illinois Urbana-Champaign, Urbana, IL, USA
}%
\affiliation{%
Holonyak Micro and Nanotechnology Laboratory, University of Illinois Urbana-Champaign, Urbana, IL, USA
}%

\date{\today}

\begin{abstract}
Intrinsic limits to device performance arise from fundamental material properties that define the best achievable operation, independent of engineering constraints. In GaN high-electron-mobility transistors (HEMTs), hot longitudinal optical (LO) phonons can act as an intrinsic performance bottleneck by reducing electron saturation velocity, output current, and transconductance--metrics that are important for device operation. While bulk GaN studies report LO phonon lifetimes of $\sim$1\,ps, leading to strong nonequilibrium phonon populations, ungated heterostructures show much shorter lifetimes of only tens of femtoseconds. Because direct measurement in HEMTs is challenging, the true impact of hot phonons remains uncertain. Full-band transport simulations of a fabricated GaN HEMT presented here reveal that LO phonon lifetimes must be $\lesssim$\,40\,fs to reproduce measured I–V characteristics, consistent with ultrafast decay observed in GaN heterostructures. We show that even this ultrafast LO-phonon decay is insufficient to fully suppress hot-phonon effects: the residual nonequilibrium LO population continues to limit the current density at high bias. Moreover, when the LO-phonon lifetime exceeds a few tens of femtoseconds, a pronounced hot-phonon bottleneck emerges, leading to a substantial current-density suppression that is inconsistent with experimental observations.
\end{abstract}

\maketitle

\section{\label{sec:intro} Introduction} 
\vspace{-10pt}
GaN-based high-electron-mobility transistor (HEMT) is an attractive technology for high-frequency and high-power electronics needed in satellite communications, radar systems, and wireless communication networks.~\cite{hoo2021emerging, lu2025review} The exceptional material properties of GaN, including its wide bandgap, high breakdown voltage,~\cite{meneghesso2014breakdown} and large peak electron velocity,~\cite{barker2005bulk} enable devices to operate under high electric fields and deliver superior power densities. The strong spontaneous and piezoelectric polarizations in AlGaN/GaN heterostructures give rise to a high-density two-dimensional electron gas (2DEG) at the interface without the need for modulation doping,~\cite{ambacher1999two} an intrinsic feature that forms the basis for the excellent electrical behavior observed in GaN HEMTs.

Electron-phonon scatterings in a device often set intrinsic limits on device performance, independent of engineering or processing constraints. In the case of GaN, the optical phonon energy is high because of the light mass of the nitrogen atom, which combined with the high electronegativity of nitrogen, results in a high electron-LO phonon scattering rate.~\cite{khurgin2007hot} If a hot electron in GaN relaxes its energy by emitting an LO phonon that does not decay rapidly into acoustic modes, it could lead to the accumulation of hot LO phonons~\cite{srivastava2008origin,barman2004long,Ridley_1996,ridley2004,ridley2013quantum} since the velocity of optical phonons is negligible.~\cite{matulionis2006hot,fang2012effect} This accumulation of hot LO phonons could result in an intrinsic performance bottleneck of the GaN technology, necessitating innovations in heat extraction and phonon engineering. This paper’s central goal is to clearly answer the question of whether hot LO phonons are critical to the performance of a GaN HEMT and compare our predictions of hot LO phonon lifetimes to prior measurements in ungated GaN heterostructures.

Electron transport in a fabricated GaN HEMT is simulated using our in-house deterministic full-band Boltzmann solver, Fermi Kinetics Transport (FKT),~\cite{10.1063/1.3270404,grupen2011energy,7505603,tunga2022comparison} which explicitly captures LO phonon dynamics. Simulations compared to measured I–V characteristics confirm that, unlike bulk GaN where LO phonon lifetimes can be relatively long, the effective lifetime in GaN HEMTs is significantly shorter.
This suppression of phonon lifetime is attributed to plasmon–phonon coupling and the influence of heterointerfaces inherent to HEMT structures.~\cite{Ridley_Dyson_2009,dyson2011lifetime}
Additional simulations show that, even with ultrafast LO-phonon decay, the hot-phonon bottleneck in GaN HEMTs persists, leading to as much as 30-60\% reduction in both saturation current and peak transconductance, reinforcing that LO phonons remain an intrinsic performance limiter. Moreover, LO-phonon lifetimes exceeding a few tens of femtoseconds produce phonon heating and I–V characteristics that deviate markedly from experimental measurements, indicating that such long lifetimes are not physically realistic for GaN HEMTs.

\begin{figure*}
    \centering
    \includegraphics[width=\linewidth]{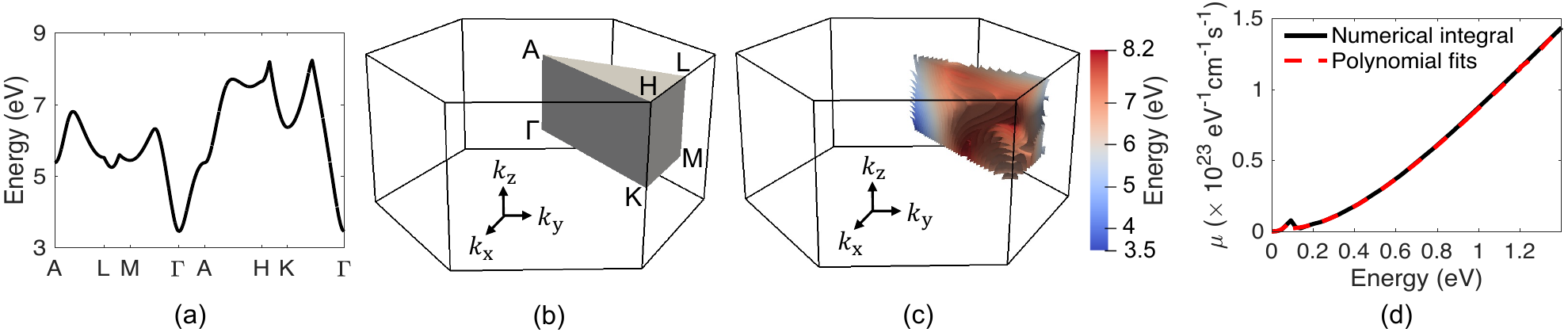}
    \vspace{-15pt}
    \caption{Workflow for preprocessing transport quantities: from (a) EPM bandstructure, (b) Brillouin-zone computation, and (c) iso-surface extraction to (d) polynomial fitting of transport integrals used in FKT simulations.}
    \label{fig:preprocess_workflow}
\end{figure*}

\section{Fermi Kinetics Transport Solver}
\vspace{-10pt}

Our work is based on FKT, which solves the Boltzmann transport equation (BTE) in the deterministic sense, making it computationally more efficient than Monte Carlo solvers. 
Unlike other deterministic Boltzmann solvers, FKT uniquely handles electronic heat flow using the heat capacity of an ideal Fermi gas instead of relying on an arbitrarily defined electron thermal conductivity. 
This alternative closure relation enhances the numerical stability and convergence of FKT compared to traditional solvers.~\cite{10.1063/1.3270404} FKT employs the method of moments to compute both electron and energy fluxes and is capable of incorporating detailed electronic bandstructures and key scattering mechanisms based on Fermi's golden rule.~\cite{grupen2011energy} 
It can also couple rapid hot electron dynamics with much slower defect ionization processes to simulate operational instabilities and with full wave electromagnetic fields to enable large signal RF simulations.~\cite{7548427} These features make FKT a versatile tool for advanced electron transport modeling.

\vspace{-10pt}
\subsection{\textbf{Incorporating Electronic Band Structure and Scattering}}
\vspace{-10pt}
Within the FKT framework, electron energy isosurfaces derived from the electronic band structure serve as the basis for precomputing all energy-dependent quantities needed for transport simulation in the device. This procedure is carried out in bulk GaN, prior to carrying out the transport simulations. The carrier density and electron particle flux are expressed in terms of isosurface integrals that depend on the electronic band structure and scattering mechanisms. For example, the carrier density is given by:
\begin{equation}
n_{i} = \int_{E_{i}}\left[\int_{\mathbf{k}_i}\underbrace{\frac{1}{4\pi^3}\frac{\delta(E_\mathbf{k} - E)}{|\nabla_{\mathbf{k}}E|}}_{\rho_\mathbf{k}(E)}d\mathbf{k}\right]f_{i}(E)dE,
\end{equation}
where $f_\mathrm{i}(E)$ is the Fermi-Dirac distribution function assigned to the $i$-th local carrier population, and the term in square brackets is the density of electron states $\rho_\mathrm{k}(E)$ integrated over all degenerate momentum vectors associated with the local distribution function $f_{\mathrm{i}}(E)$. Since multiple Fermi-Dirac distributions are assigned to a given point in real space, the total density is obtained by summing the contributions. From the first moment of the BTE, the particle flux is given by an integral of the form:
\begin{equation}
    J_\mathrm{i} = \int_{E_\mathrm{i}} 
    \underbrace{\left[ \int_{\mathbf{k}_\mathrm{i}} \mathbf{v}\,\mathbf{v}^T \, \tau_{\mathrm{k}} \, \rho_{\mathrm{k}}(E) \, d\mathbf{k} \right]}_{\mu} 
    \left( qE \, \frac{\partial f_\mathrm{i}}{\partial E} - \nabla f_\mathrm{i} \right) dE, \label{eqn:Ji}
\end{equation}
where electron velocity $\mathbf{v} = \nabla_\mathbf{k} E/\hbar$, and $\tau_\mathrm{k}(E)$ is the momentum relaxation time contributed by all scattering mechanisms, while $\mu$ represents the 
energy isosurface integral.~\cite{7505603}
Note that electrostatic screening is included in the charged impurity and piezoelectric scattering mechanisms but is neglected in the optical phonon scattering processes. 
The collision operator for polar optical phonon scattering in our simulations is derived from the Fr\"{o}hlich Hamiltonian. 
Although screening is not explicitly included in this operator at
present, we believe this to be a reasonable approximation because preliminary estimates of
the Thomas-Fermi length~\cite{go2025theory} indicate that screening may reduce our collision operator by, at
most, a factor of 4 and would likely be even less significant in the HEMT's access regions
where the collision operator has its greatest impact on device behavior. 
Nevertheless, as part of ongoing improvements to the solver, we plan to further investigate screening of polar optical phonon collision operator.

The overview of the entire preprocessing stage is illustrated in Fig.~\ref{fig:preprocess_workflow}. 
An empirical pseudopotential method (EPM)-computed band structure,~\cite{6022750} shown in Fig.~\ref{fig:preprocess_workflow}(a), was employed for wurtzite GaN, and the irreducible wedge of its first Brillouin zone, shown in Fig.~\ref{fig:preprocess_workflow}(b), was filled with a conformal tetrahedral mesh to determine the electron energies at discrete $\mathbf{k}$-points.
Using piecewise-linear interpolation along the mesh edges, the electron energy isosurfaces are extracted~\cite{grupen2011energy} as shown in Fig.~\ref{fig:preprocess_workflow}(c). The integral $\mu$ in Eq.~(\ref{eqn:Ji}) is computed over each energy isosurface to produce an energy spectrum like that shown in Fig.~\ref{fig:preprocess_workflow}(d).
Because lattice temperature, LO phonon temperature, ionized defect density, and electron density vary sensitively during device operation, the flux isosurface integrals must evolve correspondingly. In order to capture the dependence of these four quantities, the computed integrals were first fitted to polynomials of electron energy, and the resulting polynomial coefficients were subsequently parameterized as functions of the four varying quantities, \emph{i.e.}, lattice temperature, LO phonon temperature, ionized defect density, and electron density. Therefore, the simulator is capable of using these four changing quantities to obtain the correct flux isosurface integrals matching the actual numerical integrals.

\begin{figure}
    \centering
    \vspace{-10pt}
    \includegraphics[width=0.8\linewidth]{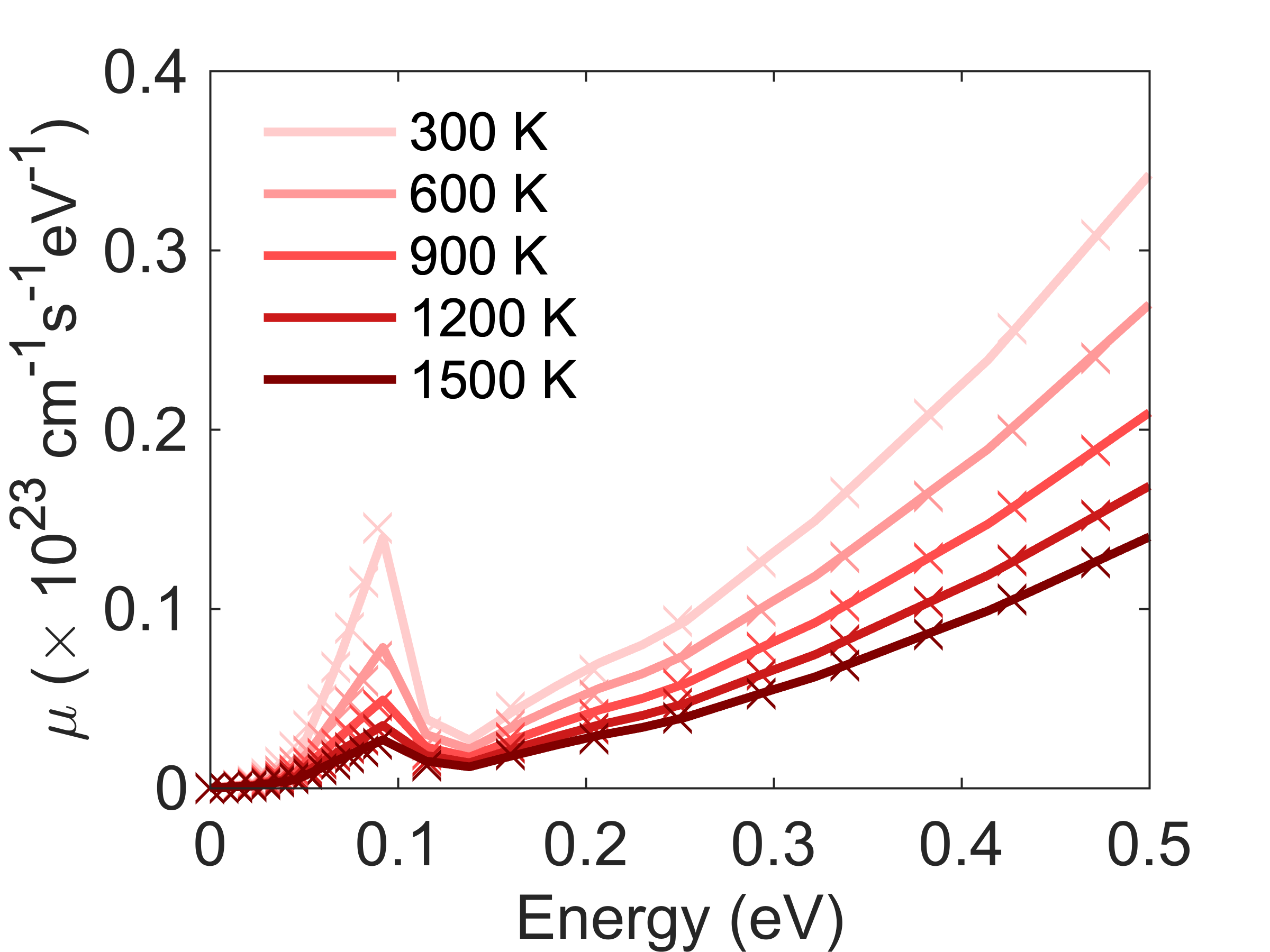}
    \caption{$\Gamma$ valley flux isosurface integrals in the horizontal plane of GaN with varying LO phonon temperatures at 300\,K lattice temperature and $\mathrm{10^{18} cm^{-3}}$ ionized defect and mobile electron densities.}
    \label{fig:flux_iso}
\end{figure}

Figure~\ref{fig:flux_iso} highlights the necessity to explicitly consider LO phonon temperature in the solver. This figure plots the $\Gamma$ valley isosurface integrals when the LO phonon temperature is varied while the other parameters (lattice temperature, ionized defect density, and electron density) are held constant. 
The energy isosurface integral increases with energy up to approximately 92 meV, driven by the growth of the velocity vector $\mathbf{v}$ and the density of states $\rho_\mathbf{k}(E)$, even though $\tau_\mathrm{k}(E)$ itself begins to decline. At around 92 meV, when electrons begin to emit optical phonons in addition to absorbing them, $\tau_\mathrm{k}(E)$ decreases sharply, leading to a reduction in the overall integral. At higher energies, the integrals gradually rise again. With increasing LO phonon temperature, the phonon occupation number rises, enhancing both absorption and emission scattering rates. Consequently, $\tau_\mathrm{k}(E)$ decreases across all electron energies, leading to a uniform reduction in the flux isosurface integrals throughout the spectrum.

\vspace{-10pt}
\subsection{Treatment of Acoustic and Optical Phonons}
\vspace{-10pt}
In the quasi-static setup employed here, the rotational electric and magnetic fields are ignored, while FKT self-consistently solves Poisson's equation, mobile electron particle and energy continuity, as well as the conservation of lattice energy. The lattice energy conservation is given as 
\begin{equation}
\rho C_\mathrm{p} \frac{\partial T_\mathrm{A}}{\partial t} + \nabla \cdot \kappa \nabla T_\mathrm{A} - C^\mathrm{E} = 0,
\label{eqn:lattice_thermal_energy}
\end{equation}
where $T_\mathrm{A}$ is the acoustic phonon (lattice) temperature, 
$C_\mathrm{p}$ is the specific heat capacity, $\kappa$ is the 
effective thermal conductivity (determined by the longitudinal acoustic (LA) and transverse acoustic (TA) modes in GaN), and  
$C^\mathrm{E}$ is the energy collision operator. When hot LO phonons are considered, $C^\mathrm{E}$ is replaced with the LO phonon's decay rate times its energy $\hbar\omega_{\rm LO}$.

\begin{figure}
    \centering
    \includegraphics[width=0.8\linewidth]{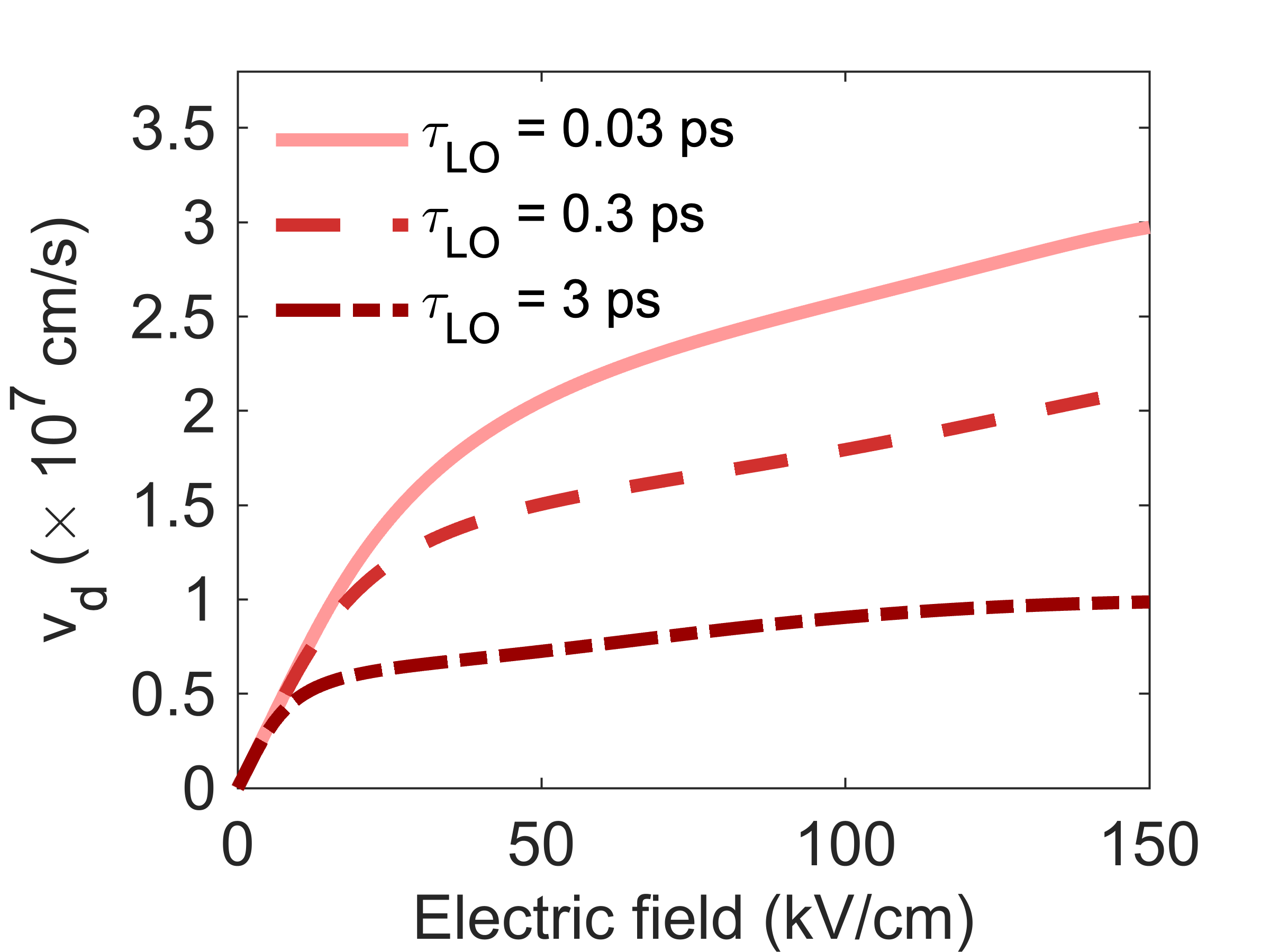}
    \vspace{-10pt}
    \caption{Bulk electron drift velocity versus field for intrinsic GaN at room temperature for different LO phonon lifetimes.}
    \label{fig:bulk_vE}
\end{figure}

\begin{figure}
    \centering
    \includegraphics[width=0.8\linewidth]{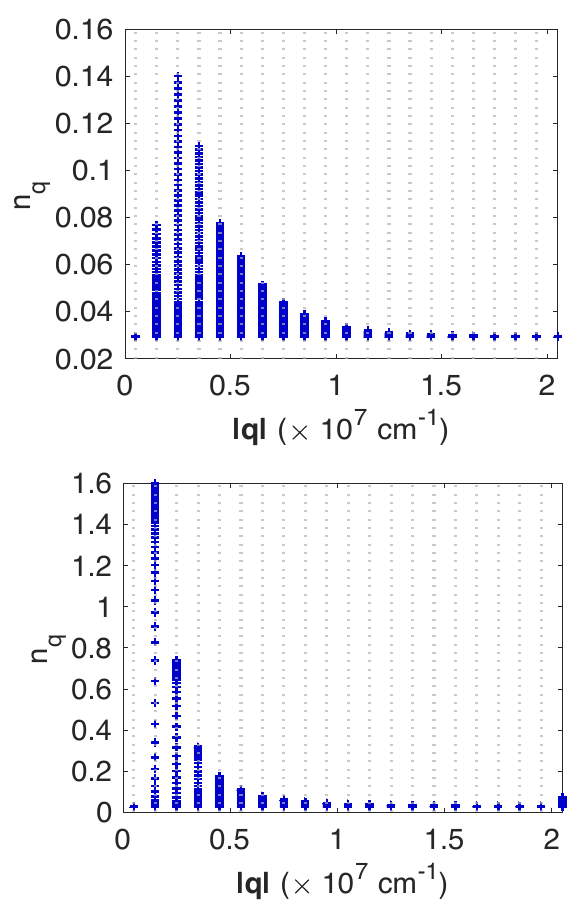}
    \vspace{-10pt}
    \caption{Occupation numbers $n_q$ for LO phonons emitted into different ranges of phonon momentum $|\mathbf{q}|$ during the simulation of electron drift velocity in bulk GaN\@. The value of $n_q$ is plotted as a symbol in each range of $|\mathbf{q}|$ for different applied fields, where $|\mathbf{E}|_{\mathrm{max}}=70$~kV/cm (top) and 600~kV/cm (bottom).
    $\tau_{\mathrm{LO}}=0.3$~ps was used for each range of $|\mathbf{q}|$.
    }
    \label{fig:nq_vs_q}
    \vspace{-10pt}
\end{figure}

Under high electric fields, the generation of LO phonons increases, which also increases the electron momentum relaxation rate due to the electron-LO scattering events. The finite lifetime of LO phonons ($\tau_\mathrm{LO}$) 
results in a non-equilibrium population of LO phonons, also termed as hot phonons. The distribution of LO phonons is given by the following mode rate equation:
\begin{equation}
\frac{1}{8\pi^3} \int d{\bf{q}} \frac{dn_\mathrm{q}}{dt}  = U_\mathrm{LO} -\frac{1}{8\pi^3} \int d {\bf{q}}\frac{n_\mathrm{q} - n_\mathrm{q}^0}{\tau_{\mathrm{LO}}},
\label{eqn:LO_decay}
\end{equation}
where $U_\mathrm{LO}$ is the LO phonon collision operator $\hbar\omega_{\rm LO}U_{\rm LO}=C^\mathrm{E}$, $n_\mathrm{q}$ is the phonon number, and $n_\mathrm{q}^0$ is the LO phonon occupation number when its temperature $T_{\rm LO}=T_{\rm A}$.
For simplicity, $n_\mathrm{q}$ in Eq.~(\ref{eqn:LO_decay}) is assumed symmetric in $\mathbf{q}$-space. To estimate $|\mathbf{q}|_{\mathrm{max}}$ for hot LO phonons, a series of bulk GaN electron drift velocity versus electric field simulations, like those shown in Fig.~\ref{fig:bulk_vE}, were conducted. As indicated in Fig.~\ref{fig:nq_vs_q}, these simulations also computed LO phonon occupation numbers for different ranges of $|\mathbf{q}|$. Based on these data, $|\mathbf{q}|_{\mathrm{max}}=6\times10^6$~1/cm was used in Eq.~(\ref{eqn:LO_decay}) for GaN HEMT device simulations.


\vspace{-10pt}
\section{GaN HEMT Setup and Simulation Results}
\vspace{-10pt}
\subsection{\textbf{Device Configuration}}
\vspace{-10pt}

\begin{figure}[h!]
    \centering
    \includegraphics[width=0.75\linewidth]{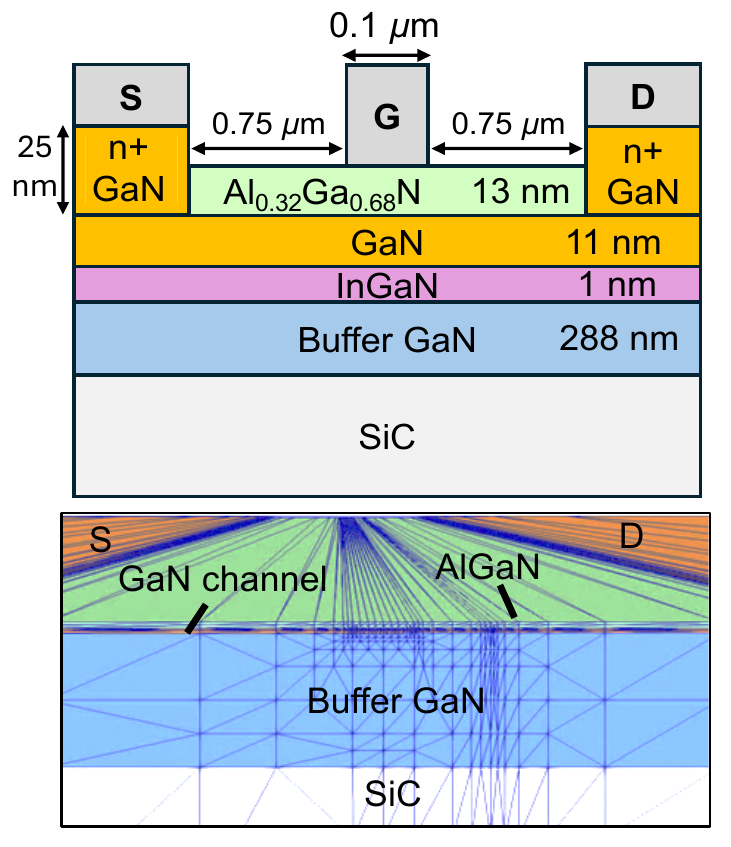}
    \vspace{-10pt}
    \caption{Cross-section of the fabricated GaN HEMT (top), and the meshed structure (bottom).}
    \label{fig:Device_str}
    \vspace{-15pt}
\end{figure}

The device structure is illustrated in Fig.~\ref{fig:Device_str}. The heterostructure layers were grown by metalorganic chemical vapor deposition (MOCVD) with growth temperatures of various layers adjusted to promote good incorporation, defect reduction, and thickness control.~\cite{1637547} 
The device layers are grown on a SiC substrate for improved thermal conduction. A 1\,nm $\mathrm{In}_{0.1}\mathrm{Ga}_{0.9}\mathrm{N}$ layer is first grown on GaN buffer on SiC and followed by an 11\,nm GaN channel. The top barrier is a 25\,nm $\mathrm{Al}_{0.32}\mathrm{Ga}_{0.64}\mathrm{N}$ layer, which is recessed by 12\,nm, leading to a 13\,nm thick layer underneath the gate contact. 
The back barrier was shown to promote electron confinement and superior gate modulation efficiency. 
The gate length is 0.1~$\mu$m, with source-to-gate (gate-to-drain) spacing of 0.75~$\mu$m. Additional device-specific parameters are listed in Table \ref{table:matl_param}.

The device simulation employs a Dirichlet boundary condition at the bottom of the substrate that holds the lattice temperature equal to 300\,K, while a Neumann boundary condition is used at other surfaces. Perfect electric conductors (PECs) on top of the device are treated as iso-thermal bodies. 

\begin{table}[h!]
  \caption{Device parameters.}
  \label{tab:params}
  \begin{ruledtabular}
  \begin{tabular}{l c}
    \textrm{\bf{Parameter}} & \textrm{\bf{Value}} \\
    \colrule
    \\
    Source/Drain doping                    & $2\times10^{19}\ \mathrm{cm^{-3}}$ \\
    Gate Schottky barrier height           & $1.2\ \mathrm{eV}$ \\
    Polarization sheet charge density      & $1.115\times10^{13}\ \mathrm{cm^{-2}}$ \\
    Surface trap density (source access)   & $1.3-3.5\times10^{12}\ \mathrm{cm^{-2}}$ \\
    Surface trap density (drain access)    & $5.2\times10^{12}\ \mathrm{cm^{-2}}$ \\
    LA/TA phonon deformation potential     & $8.3\ \mathrm{eV}$ \\
    LO phonon deformation potential        & $10^{9}\ \mathrm{eV/cm}$ \\
    LO phonon energy                       & $92\ \mathrm{meV}$ \\
  \end{tabular}
  \label{table:matl_param}
  \end{ruledtabular}
\end{table}

\vspace{-10pt}
\subsection{\textbf{Device Simulation} \label{sec:dev_sim}}
\vspace{-10pt}

\begin{figure}[h!]
    \centering
    \includegraphics[width=0.8\linewidth]{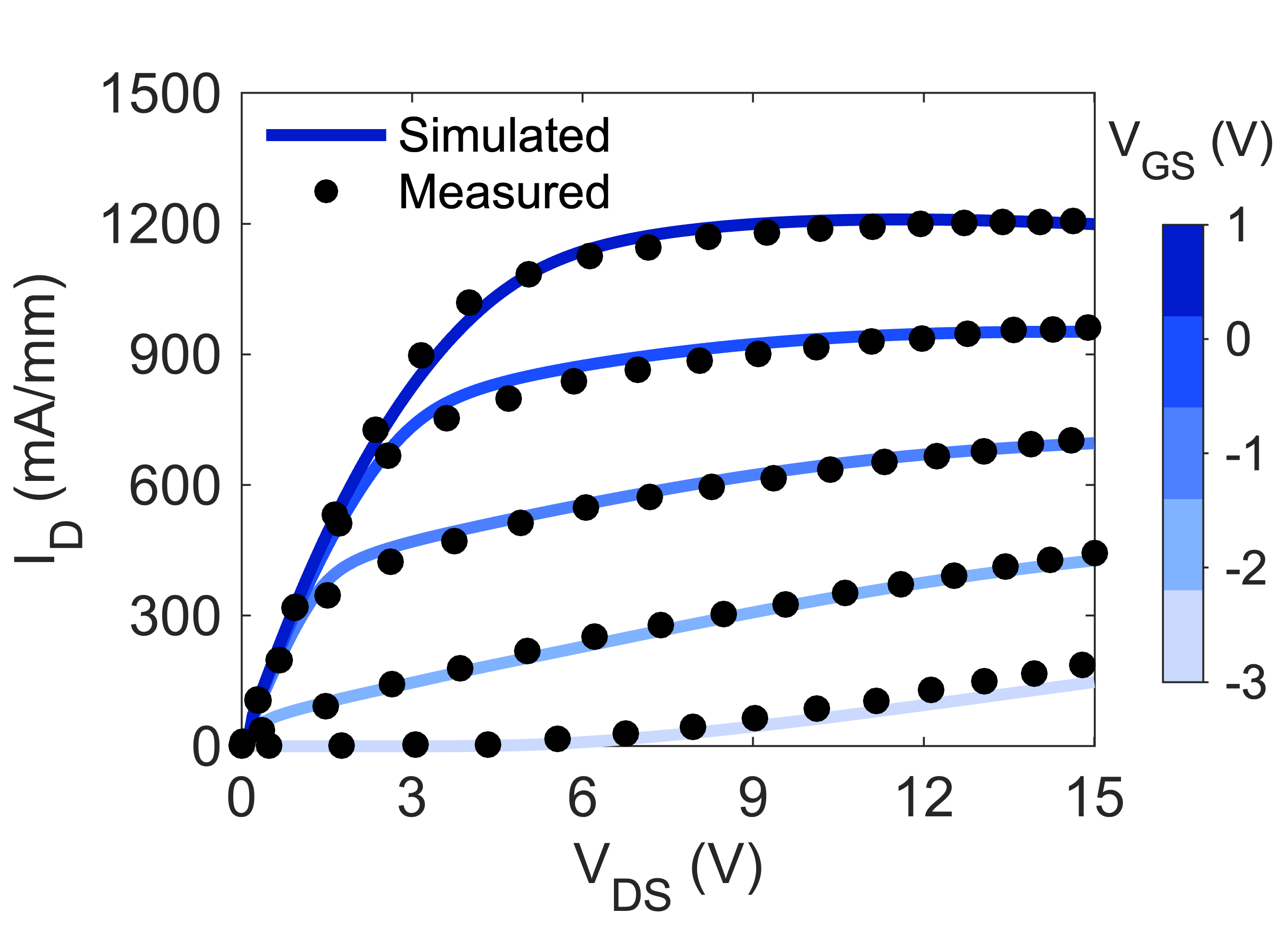}
    \vspace{-15pt}
    \caption{Simulated output curves with 30 fs hot LO phonon lifetime. Symbols are experimental data from Marino \emph{et al.}~\cite{Marino2010}}
    \label{fig:I-V}
\end{figure}

\begin{figure}[h]
    \centering
    \includegraphics[width=0.8\linewidth]{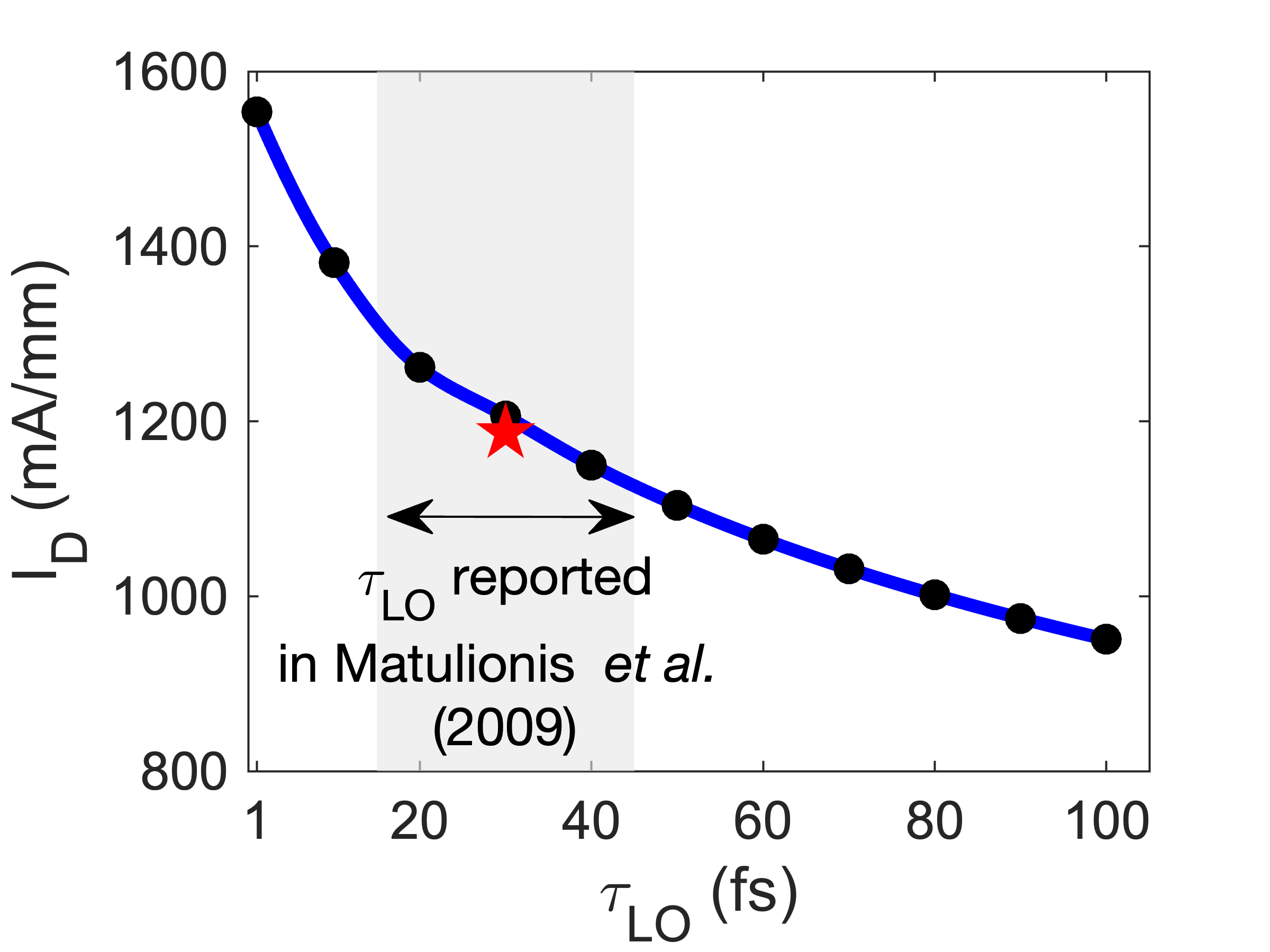}
    \vspace{-10pt}
    \caption{Degradation of steady-state current at a fixed bias ($V_\mathrm{GS}$ = 1\,V, $V_\mathrm{DS}$ = 10\,V) with increasing LO phonon lifetime. The value of $\tau_\mathrm{LO}$ used to compute the currents in Fig.~\ref{fig:I-V} is marked in red.}
    \label{fig:id_tau}
    \vspace{-10pt}
\end{figure}

Measured and simulated GaN HEMT drain currents $I_\mathrm{D}$ versus drain voltage $V_\mathrm{DS}$ for different gate biases $V_\mathrm{GS}$ are shown in Fig.~\ref{fig:I-V}. The favorable comparison between measured and simulated data indicates that FKT can accurately capture the physics of the fabricated device.~\cite{Marino2010} The solver predicts that the LO phonon lifetime must be as short as few tens of fs in order to explain the measured I–V characteristics. Specifically, for the results in Fig.~\ref{fig:I-V}, $\tau_\mathrm{LO} = 30$\,fs. 
While such a short phonon lifetime may seem unexpected for bulk GaN, it is reasonable in GaN heterostructures, where LO phonon interactions with interfaces create additional decay channels that enable rapid conversion into heat-conducting modes.~\cite{matulionis2009ultrafast,liberis2014hot} 
The impact of $\tau_\mathrm{LO}$ on the output current at $V_\mathrm{GS}=1$\,V and $V_\mathrm{DS}=10$\,V is shown in Fig.~\ref{fig:id_tau}. 
As expected, increasing $\tau_\mathrm{LO}$ beyond 30\,fs reduces the drain current. Conversely, decreasing $\tau_\mathrm{LO}$ from 30 fs to 1 fs yields a $\sim$30\% increase in drain current, indicating that even at ultrashort $\tau_\mathrm{LO}$ the device performance remains constrained by the hot-LO-phonon bottleneck.

\begin{figure*}
    \centering
    \includegraphics[width=0.8\linewidth]{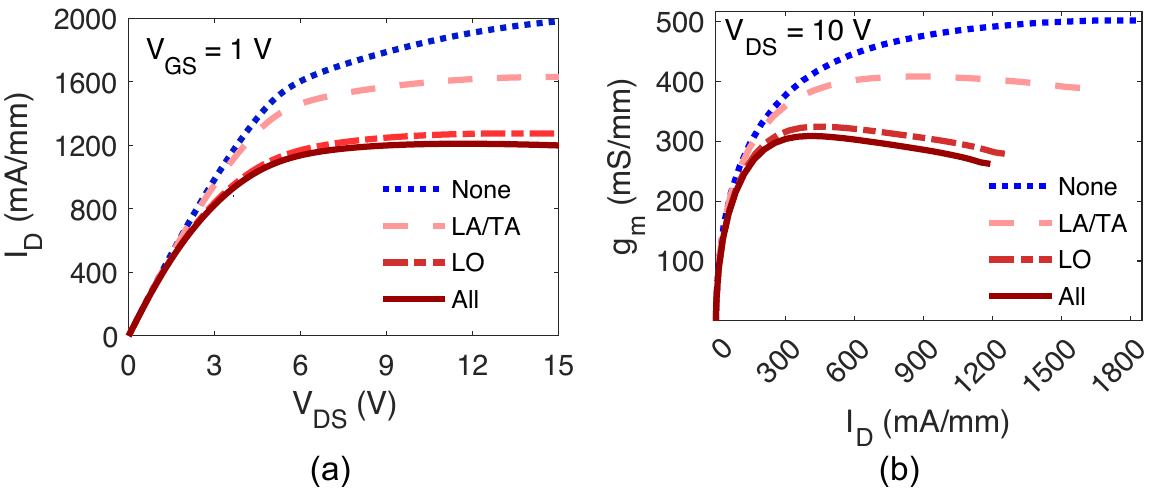}
    \vspace{-10pt}
    \caption{(a) Simulated steady-state current at a fixed gate bias showing the impact of LO phonon heating. The inclusion of LO phonon heating significantly reduces the current compared to models considering only acoustic phonon heating (LA/TA) or neglecting phonon heating altogether. Even in the absence of acoustic phonon heating, the LO phonon contribution alone is sufficient to notably degrade the output current. 
    (b) Impact of phonon heating on transconductance ($g_\mathrm{m}$) at $V_\mathrm{DS}$ = 10\,V. }
    \label{fig:Id_vg1_gm}
\end{figure*}

The role of LO phonon heating is further illustrated in Fig.~\ref{fig:Id_vg1_gm}(a), which presents the simulated output characteristics at $V_{\mathrm{GS}} = 1~\mathrm{V}$ under four different conditions. The simulation with $\tau_{\mathrm{LO}} = 30$~fs is the baseline result (labeled as ``All''), which matches well with the measured data shown in Fig.~\ref{fig:I-V}. When LO phonon heating with $\tau_{\mathrm{LO}} = 30$~fs is considered but acoustic phonon heating is ignored (result labeled as ``LO''), the drain current merely increases by 5\% at $V_\mathrm{GS} =1$\,V and $V_\mathrm{DS}=10$\,V, compared to the baseline case. This observation reinforces that LO phonon heating alone introduces substantial performance degradation even in the absence of acoustic-phonon-induced self-heating.
When we assume an instantaneous decay of LO phonons into acoustic modes, effectively removing the hot-LO phonon contribution and considering only longitudinal and transverse acoustic phonon heating (labeled as ``LA/TA''), $I_{\mathrm{D}}$ increases by approximately 30\% at $V_\mathrm{GS}=1$\,V and $V_\mathrm{DS}=10$\,V.
In contrast, 
eliminating both hot LO and hot acoustic phonons in the device (labeled as ``None'') would increase the device current by up to 50\%, suggesting that phonon-engineering strategies could, in principle, enhance intrinsic device performance. However, we are not aware of any studies demonstrating LO-phonon lifetimes below $30\pm 15$\,fs in GaN heterostructures, indicating that such aggressive lifetime reduction has not yet been realized experimentally.

Similar to the trends observed for $I_\mathrm{D}$, the transconductance, $g_\mathrm{m}$, is also significantly affected by the degree of LO-phonon heating, even when hot phonons decay rapidly into acoustic modes. As shown in Fig.~\ref{fig:Id_vg1_gm}(b), 
the 
peak $g_\mathrm{m}$ would increase by roughly 30\% 
in the absence of LO-phonon heating, and by nearly 60\% if all phonon heating were eliminated.

\begin{figure*}
    \centering
    \includegraphics[width=0.8\linewidth]{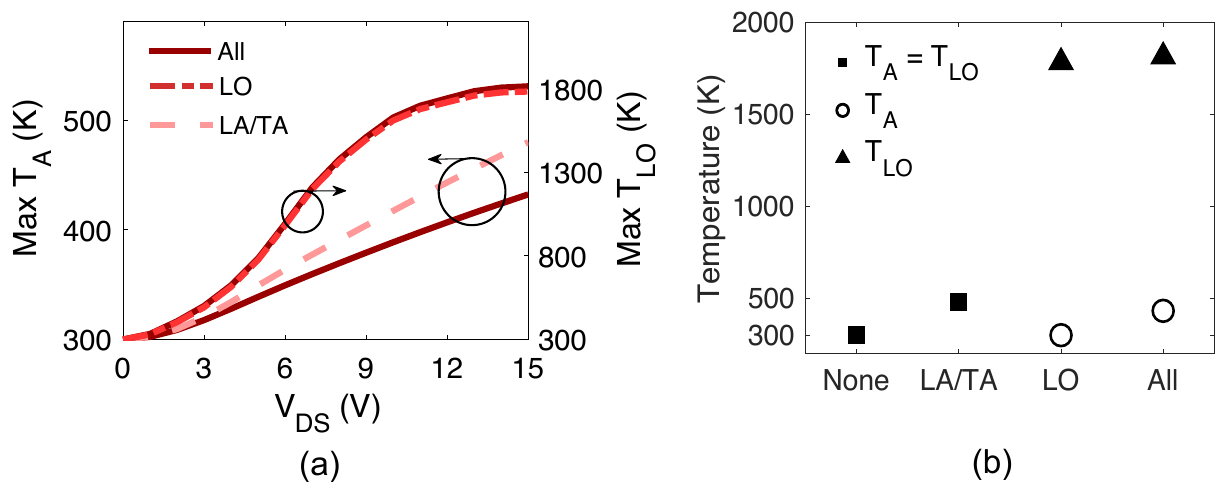}
    \vspace{-12pt}
    \caption{(a) Gradual rise of peak acoustic phonon temperature and LO phonon temperature in the channel versus $V_\mathrm{DS}$ at $V_\mathrm{GS}$ = 1\,V. (b) Peak acoustic phonon temperature and LO phonon temperature in the channel ($V_\mathrm{GS}$ = 1\,V, $V_\mathrm{DS}$ = 15\,V). For cases with LO phonon heating, $\tau_\mathrm{LO}=30\,$fs.
    }
    \label{fig:max_temp_4cases}
\end{figure*}

\vspace{-10pt}
\subsection{\textbf{Electron and Phonon Temperatures}}
\vspace{-15pt}

Examining electron and phonon temperatures within the GaN HEMT may offer insights into the device behaviors presented in Section~\ref{sec:dev_sim}. Figure~\ref{fig:max_temp_4cases}(a) shows the peak phonon temperature in the GaN channel for the simulation configurations corresponding to Fig.~\ref{fig:Id_vg1_gm}(a).
When only acoustic phonon heating is included, 
$T_{\mathrm{A}}$ rises uniformly above 300\,K due to Joule heating. Since LO phonon heating is not considered in this case, the temperature of optical phonons is the same as that of acoustic phonons.
However, when hot-phonon effects are explicitly included with $\tau_{\mathrm{LO}} = 30$\,fs, a clear decoupling between $T_{\mathrm{A}}$ and $T_{\mathrm{LO}}$ emerges. The LO phonon temperature increases sharply with drain bias, exceeding 1800\,K at high fields, while the acoustic phonon temperature remains substantially lower. 
Figure \ref{fig:max_temp_4cases}(b) further highlights how the assumed heating mechanisms lead to distinct phonon temperature profiles, particularly under the high drain bias where these differences are most pronounced.

\begin{figure}[h!]
\vspace{-10pt}
    \centering
    \includegraphics[width=0.75\linewidth]{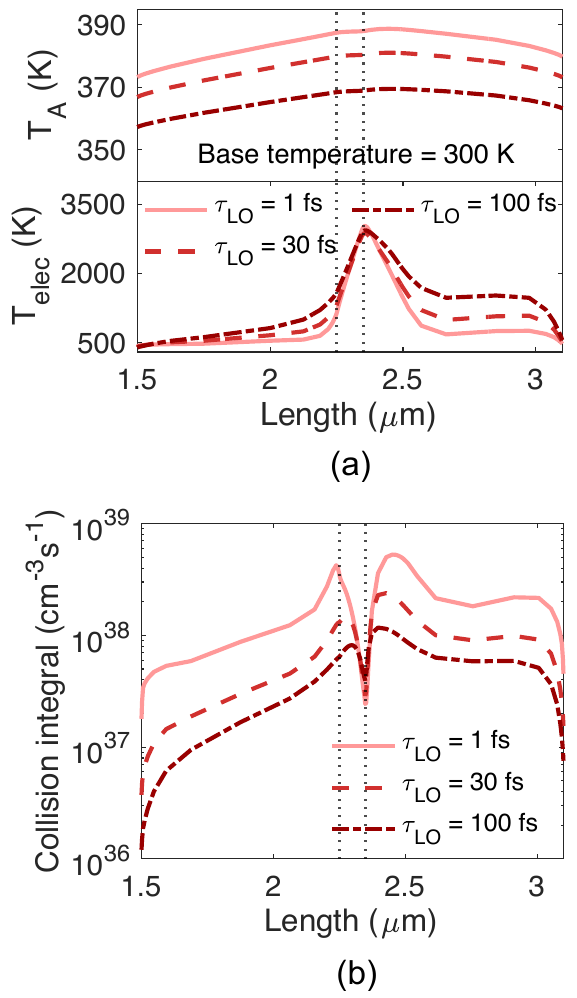}
    \vspace{-10pt}
   \caption{(a) Acoustic phonon temperature (top) and electron temperature (bottom), and (b) collision integral along the channel length at 0.1 nm below the AlGaN/GaN interface ($V_\mathrm{GS}$ = 1\,V, $V_\mathrm{DS}$ = 10\,V). The channel is marked by the dotted vertical lines.}
    \label{fig:Ta_Te}
\end{figure}

We also investigate the spatial profiles of lattice and electron temperatures along the channel length probed directly below the AlGaN/GaN interface, as shown in Fig.~\ref{fig:Ta_Te}(a). Although the lattice heats only up to about 390\,K at $V_\mathrm{GS}=1$\,V and $V_\mathrm{DS}=10$\,V, the electron temperatures can be as high as 2800\,K. 
Furthermore, we can see that the electron temperature peaks sharply near the gate–drain edge. This region coincides with the high electric field in the channel, where carriers gain substantial kinetic energy and predominantly lose it through LO phonon emission. Although the electron temperature is elevated at the gate–drain edge, the carrier density in this pinch-off region is substantially depleted, and this low density of hot electrons traverses the pinch-off region at a high rate compared to the phonon scattering rate. As a result, the number of LO phonons emitted by these hot electrons is moderate, as shown in Fig.~\ref{fig:Ta_Te}(b), causing the nonequilibrium LO-phonon occupation number to stay close to its equilibrium value within this narrow depleted region.
Just beyond the pinch-off region, however, in the access region, the electron density increases rapidly, and access region electrons heat up by absorbing the energy carried by the nearly ballistic electrons that emerge from the pinch-off region. This results in a much larger population of carriers in the access region capable of emitting LO phonons.
This enhanced emission drives the LO-phonon occupation above equilibrium, particularly for longer LO-phonon lifetimes, which increases phonon reabsorption and reduces electron cooling. As a result, we observe larger $\tau_\mathrm{LO}$ values producing elevated electron temperatures in the access regions, despite reduced net emission in the pinch-off region.

\begin{figure}[t!]
\vspace{-10pt}
    \centering
    \includegraphics[width=0.85\linewidth]{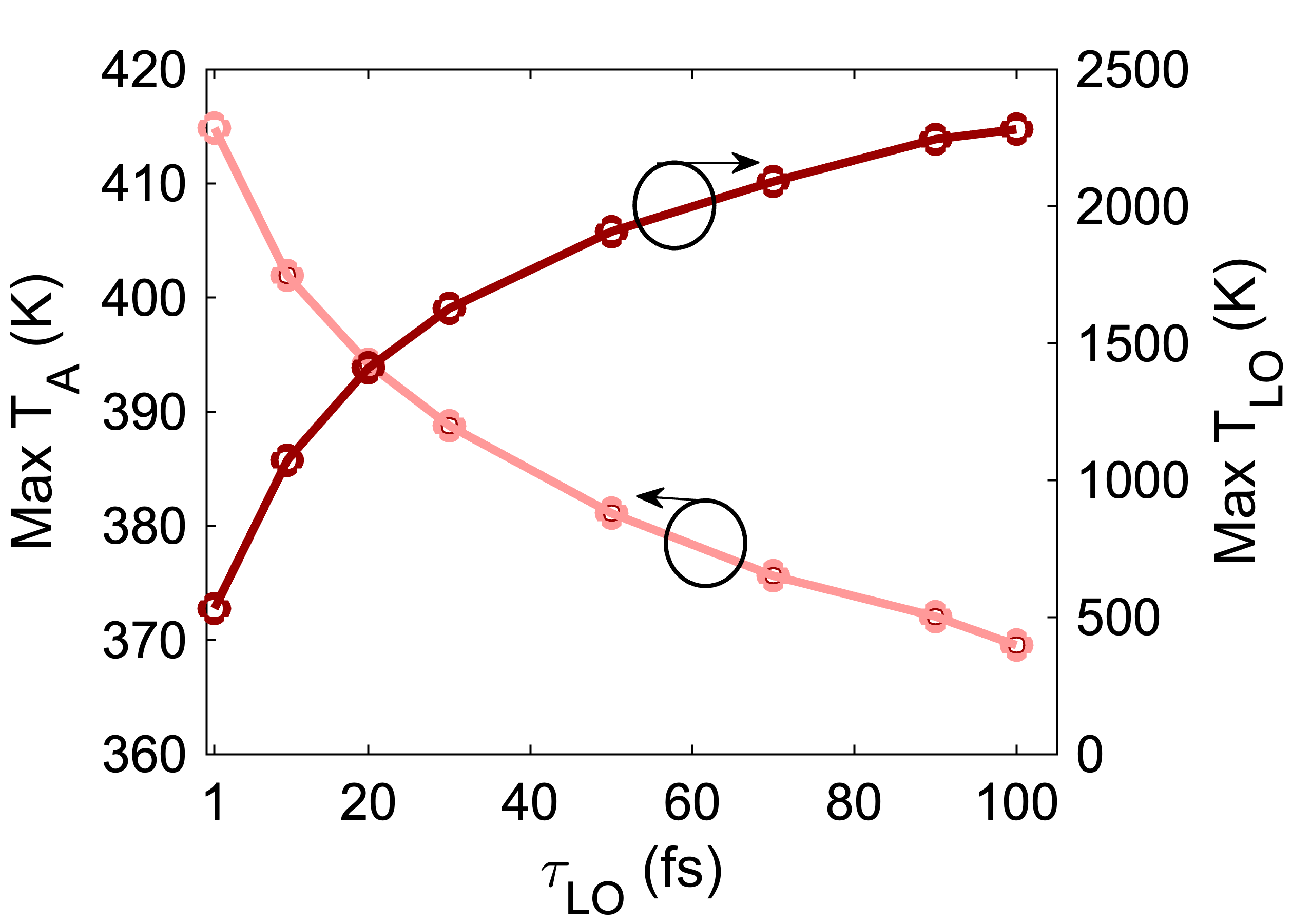}
    \vspace{-10pt}
    \caption{Peak acoustic phonon temperature and LO phonon temperature in the channel ($V_\mathrm{GS}$ = 1\,V, $V_\mathrm{DS}$ = 10\,V).}
    \label{fig:Max_temp}
\end{figure}

The temperature variations in Fig.~\ref{fig:Ta_Te}(a) also reveal certain characteristics of acoustic phonon lattice heating. In contrast to the electron temperature profile, the lattice temperature exhibits a more gradual spatial variation, increasing smoothly along the channel due to the slower energy transfer from electrons to acoustic phonons mediated by LO phonon decay. The lattice temperature's dependence on $\tau_{\mathrm{LO}}$ also appears to be in contrast to the electron temperature.
Under dynamic equilibrium conditions the power supplies inject energy into the device as $V_\mathrm{DS}I_\mathrm{D}$, which first heats the mobile electrons through Joule heating. The hot electrons, in turn, transfer their energy to the LO phonons through scattering, which subsequently decay into acoustic phonons that ultimately diffuse heat into the heat sinks. Increasing the LO phonon lifetime, $\tau_\mathrm{LO}$, strengthens the hot-phonon bottleneck where the long-lived LO phonons accumulate in the channel, forming a nonequilibrium population that repeatedly reabsorbs energy from the hot electrons. This elevates $T_\mathrm{LO}$, suppresses electron mobility, and reduces the overall current and power dissipation $V_\mathrm{DS}I_\mathrm{D}$. The resulting drop in $V_\mathrm{DS}I_\mathrm{D}$ leads to a lower rate of energy that needs to be carried away by the acoustic phonons, thereby reducing the maximum $T_\mathrm{A}$. 
This establishes a negative feedback mechanism, where we observe that increased $\tau_\mathrm{LO}$ leads to stronger electron heating but weaker lattice self-heating across the GaN channel. This explains the opposing trends of the peak lattice temperature and LO phonon temperature observed in Fig.~\ref{fig:Max_temp}, where increasing $\tau_\mathrm{LO}$ causes the peak $T_\mathrm{LO}$ to rise while the peak $T_\mathrm{A}$ decreases. Therefore, reduced lattice self-heating with longer phonon lifetime originates not directly from phonon accumulation, but from the diminished Joule heating due to lower electron mobility and current flow.

\vspace{-10pt}
\section{Discussion}
\vspace{-10pt}
In GaN transistors operating under high electric fields, electron–phonon scatterings can have a significant impact on electron transport. When nonequilibrium LO phonons are generated, they can introduce an additional intrinsic bottleneck that limits performance. Establishing benchmarks for the impact of hot LO phonons on output current and transconductance is therefore critical for understanding whether they significantly hinder further improvements in key HEMT metrics such as cutoff frequency and RF linearity, both of which are closely tied to the device transconductance. Consequently, any degradation of transconductance due to LO phonons directly constrains the RF performance of the device.

The lifetime of hot LO phonons determines the severity of their influence on device behavior. This lifetime arises from a complex interplay of factors, including ambient temperature, carrier density, and supplied power determining electron temperature.~\cite{tsen1998time,tsen2006subpicosecond,Dyson_2009a,matulionis2009ultrafast,matulionis2009hot} The lifetime is known to decrease with increasing carrier density due to plasmon–LO phonon coupling, which enhances lattice anharmonicity.~\cite{dyson2008phonon} The resulting coupled phonon–plasmon modes possess higher group velocities, allowing phonons to escape from the active channel region more efficiently. In bulk GaN, the lifetime of the coupled LO-phonon-plasmon mode is several 100s of femtoseconds, while at low carrier densities the LO phonon mode can live for a few picoseconds. In contrast, in AlGaN/GaN the involved wavevectors in electron-phonon interaction result in Landau damped plasma modes, making quantitative assessment of the lifetime more challenging.~\cite{Ridley_Dyson_2009,dyson2011lifetime}

\begin{figure}[h!]
    \centering
    \includegraphics[width=1\linewidth]{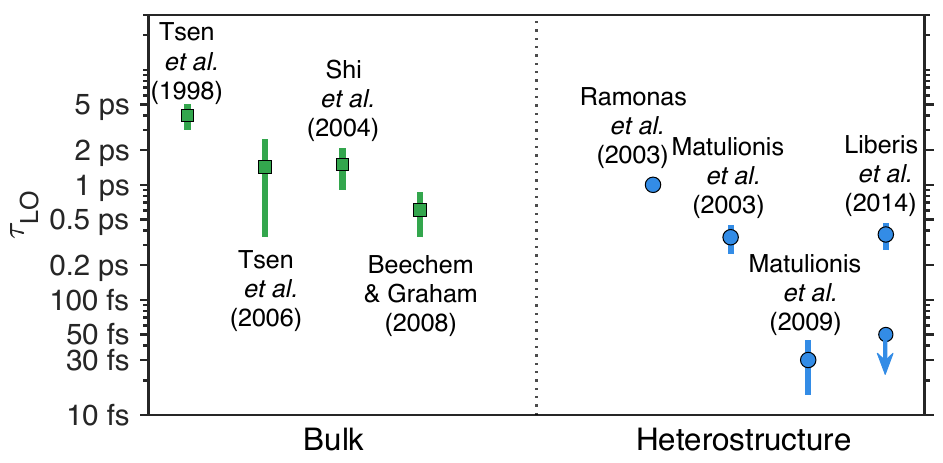}
    \vspace{-15pt}
   \caption{LO phonon lifetime reported in previous works.~ \cite{tsen1998time,tsen2006subpicosecond,shi2004raman,beechem2008temperature,ramonas2003monte,PhysRevB.68.035338,matulionis2009plasmon,liberis2014hot}}
    \label{fig:tau_LO_reference}
\end{figure}

Experimental studies of an AlGaN/GaN heterostructure by Matulionis \emph{et al.} reported lifetimes of $30\pm 15$\,ps at supplied powers of $20\pm 10$\,nW per electron, while Liberis \emph{et al.} observed lifetimes $\leq 50$\,fs for powers $\geq$10\,nW per electron. These results suggest that additional mechanisms in heterostructures can suppress the LO phonon lifetime by nearly two orders of magnitude compared to bulk GaN. 
It has been proposed that coupled phonon–plasmon modes in heterostructures can propagate toward the interface more rapidly than they decay within the channel, resulting in the observed ultrafast lifetimes.
Figure~\ref{fig:tau_LO_reference} summarizes the range of $\tau_\mathrm{LO}$ values reported in the literature for bulk GaN and AlGaN/GaN heterostructure.

FKT simulations suggest that only ultrashort LO phonon lifetimes, below 40 fs, can reproduce the measured device characteristics. Longer lifetimes lead to a pronounced hot-phonon bottleneck that contradicts experimental observations.
A key question is whether the measured DC characteristics could be reproduced by adjusting other device parameters in a way that would permit longer LO phonon lifetimes (\emph{e.g.}, $\tau_{\mathrm{LO}} \sim 100$~fs). While certain device features, such as the surface-state densities in the source and drain access regions, can be varied within physically reasonable ranges and affect the current levels to achieve a reasonable match for $\tau_{\mathrm{LO}}$ below 40~fs, the realistic variability of other model parameters is rather limited due to the fully physical basis of FKT. 
Device features such as the Schottky barrier height, polarization sheet charge density, or channel doping are tightly constrained by threshold voltage, contact resistance, and known epitaxial profiles; significant modification of these values leads to unphysical threshold voltages or incorrect subthreshold slopes. Among the device features that can be reasonably modified, none are capable of reproducing the measured I-V characteristics over the entire bias space when $\tau_{\mathrm{LO}}$ exceeds $\sim 40$~fs. Our prior simulations further demonstrate that increases in $\tau_{\mathrm{LO}}$ and other combinations of device features cannot simultaneously yield the near-threshold conduction and high-field current saturation. In contrast, ultrafast LO phonon decay ($\tau_{\mathrm{LO}} \lesssim 40$~fs) consistently yields agreement across all operating regions, irrespective of small variations in pertinent device features, \emph{e.g.} access-region surface traps.

Thus, although 
uncertainties about measured device features introduce some flexibility in defining the simulated structure, the overarching conclusion remains robust: the measured characteristics of this fabricated HEMT can only be explained when LO-phonon heating is limited by ultrashort phonon lifetimes far below those of bulk GaN. 
This outcome aligns with prior measurements in GaN heterostructures but represents the first theoretical investigation of LO-phonon lifetimes within a realistic HEMT using a full-band transport solver. The limited and physically constrained degrees of freedom available in the device model prevent long-lived LO phonons from being accommodated, reinforcing the conclusion that LO phonon lifetime must be short in practical GaN HEMTs compared to their bulk structures, yet these shortlived optical phonons can significantly suppress current density and transconductance.

We note that quantum confinement effects are not included in our simulations. This choice is motivated by the nature of carrier transport in these devices, which is dominated by hot electrons in and around the pinch-off region. In this region, the electric field perpendicular to the channel–barrier interface is reduced, resulting in weaker quantum confinement. Moreover, very hot electrons, such as those present near pinch-off, can more readily access the full three-dimensional momentum space during scattering events. Under these conditions, treating channel electrons using a three-dimensional description constitutes a reasonable approximation. Using this description, FKT has successfully captured the physics underlying the experimentally observed kink effect~\cite{grupen2019reproducing} and gate lag~\cite{miller2022experimentally} in GaN HEMTs. Moreover, most ensemble Monte Carlo–based transport simulations of GaN HEMTs rely on a similar approximation, without explicitly resolving carrier wave functions from Schr\"{o}dinger's equation.~\cite{fang2019electron, garcia2025electrothermal}

Finally, we briefly comment on our treatment of optical phonons. We assume a $\mathbf{q}$-symmetric optical phonon distribution, with the upper bound on $|\mathbf{q}|_\mathrm{max}$ determined from bulk GaN simulations, as an effective description of hot-phonon accumulation. Varying $|\mathbf{q}|_\mathrm{max}$ within a reasonable range $(5-7)\times 10^6$\,cm$^{-1}$ results in only minor quantitative differences and does not alter the main conclusions of this work.
In addition, within our solver all hot electrons are allowed to emit phonons: for a hot electron with a given initial wave vector $\mathbf{k}$ all possible final states $\mathbf{k}'$ are considered. We find that most LO phonons are emitted by electrons in the drain access region adjacent to the pinch-off region. Owing to the high electron density in this region, the asymmetry in the electron distribution is expected to be relatively small, suggesting only a weak preference for forward versus backward optical phonon emission. Consequently, while anisotropic phonon distributions may introduce quantitative corrections, the primary conclusions of the paper remain robust.

\vspace{-15pt}
\section{Conclusions}
\vspace{-10pt}

The FKT framework is employed to model hot-phonon effects in AlGaN/GaN HEMTs, incorporating full-band electronic structure and non-equilibrium LO phonon dynamics. 
The simulations showed not only that ultrashort LO-phonon lifetimes of $\lesssim$~40\,fs are required to reproduce measured DC characteristics, consistent with prior studies on GaN heterostructures, but also that, although the LO-phonon decay is fast, it is still not sufficiently fast to eliminate the hot-phonon bottleneck. As a result, the saturated output current and peak transconductance remain degraded by approximately 30\% and 60\%, respectively.
This study also highlighted the intricate interplay between electron transport, phonon dynamics, and power dissipation, demonstrating how increasing $\tau_\mathrm{LO}$ lowered carrier mobility and drain current, thereby reducing the injected power $V_\mathrm{DS}I_\mathrm{D}$ and the corresponding acoustic phonon heat flow into the sinks. Consequently, the maximum lattice temperature $T_\mathrm{A}$ decreased. 
These findings are significant for elucidating the intrinsic performance limits imposed by phonon dynamics and for guiding material and device engineering strategies aimed at achieving higher output current and transconductance.

\vspace{-10pt}
\begin{acknowledgments}
\vspace{-10pt}
This work was supported by AFOSR Grant No. LRIR 24RYCOR009, DARPA Agreement No. HR00112390072, and NSF Grant No. ECCS-2237663.
\end{acknowledgments}

\vspace{-10pt}
\section*{Author Declarations}
\vspace{-10pt}
\subsection*{Conflict of Interest}
\vspace{-10pt}
The authors have no conflicts to disclose.
\vspace{-10pt}
\subsection*{Author Contributions}
\vspace{-10pt}
\textbf{Ankan Ghosh Dastider:} Formal analysis (lead); Investigation (equal); Validation (lead); Visualization (lead); Writing original draft (equal). \textbf{Matt Grupen:} Conceptualization (equal); Investigation (equal); Software (lead); Methodology (lead); Writing original draft (equal); Resources (equal). \textbf{Ashwin Tunga:} Validation (supporting); Visualization (supporting); \textbf{Shaloo Rakheja:} Conceptualization (equal); Funding acquisition (lead); Project administration (lead); Resources (equal); Supervision (lead); Writing original draft (equal).

\vspace{-10pt}
\section*{Data Availability Statement}
\vspace{-10pt}
The data that support the findings of this study are available from the corresponding author upon reasonable request.

\appendix

\section*{References}

\bibliography{aipsamp}

@PREAMBLE{
 "\providecommand{\noopsort}[1]{}" 
 # "\providecommand{\singleletter}[1]{#1}%" 
}

@article{garcia2025electrothermal,
  title={Electrothermal modeling of {GaN} high electron mobility transistors using a {Monte Carlo-trained hybrid AI}-thermal approach with microscopic physical insight},
  author={Garc{\'\i}a-S{\'a}nchez, S and {\'I}{\~n}iguez-de-la-Torre, I and Mateos, J and Gonz{\'a}lez, T},
  journal={Journal of Applied Physics},
  volume={138},
  number={9},
  year={2025},
  publisher={AIP Publishing}
}

@article{fang2019electron,
  title={Electron Transport Properties of {Al$_\mathrm{x}$Ga$_{1-\mathrm{x}}$N/GaN} Transistors Based on First-Principles Calculations and {Boltzmann}-Equation {Monte Carlo} Simulations},
  author={Fang, Jingtian and Fischetti, Massimo V and Schrimpf, Ronald D and Reed, Robert A and Bellotti, Enrico and Pantelides, Sokrates T},
  journal={Physical Review Applied},
  volume={11},
  number={4},
  pages={044045},
  year={2019},
  publisher={APS}
}

@article{go2025theory,
  title={Theory of quasistatically screened electron-polar optical phonon scattering},
  author={Go, Yuji and Dutt, Rajeev and Neophytou, Neophytos},
  journal={Physical Review B},
  volume={111},
  number={19},
  pages={195211},
  year={2025},
  publisher={APS}
}

@article{hoo2021emerging,
  title={Emerging {G}a{N} technologies for power, {RF}, digital, and quantum computing applications: {R}ecent advances and prospects},
  author={Hoo Teo, Koon and Zhang, Yuhao and Chowdhury, Nadim and Rakheja, Shaloo and Ma, Rui and Xie, Qingyun and Yagyu, Eiji and Yamanaka, Koji and Li, Kexin and Palacios, Tomás},
  journal={Journal of Applied Physics},
  volume={130},
  number={16},
  year={2021},
  publisher={AIP Publishing}
}

@article{shi2004raman,
    author = {Shi, L. and Ponce, F. A. and Menéndez, J.},
    title = {Raman line shape of the {A}1 longitudinal optical phonon in {G}a{N}},
    journal = {Applied Physics Letters},
    volume = {84},
    number = {18},
    pages = {3471-3473},
    year = {2004},
    month = {05},
    issn = {0003-6951},
    doi = {10.1063/1.1737792}
}

@article{beechem2008temperature,
    author = {Beechem, Thomas and Graham, Samuel},
    title = {Temperature and doping dependence of phonon lifetimes and decay pathways in {G}a{N}},
    journal = {Journal of Applied Physics},
    volume = {103},
    number = {9},
    pages = {093507},
    year = {2008},
    month = {05},
    issn = {0021-8979},
    doi = {10.1063/1.2912819},
    url = {https://doi.org/10.1063/1.2912819},}

@article{ramonas2003monte,
doi = {10.1088/0268-1242/18/2/310},
url = {https://doi.org/10.1088/0268-1242/18/2/310},
year = {2003},
month = {jan},
publisher = {},
volume = {18},
number = {2},
pages = {118},
author = {M Ramonas and A Matulionis and L Rota},
title = {Monte Carlo simulation of hot-phonon and degeneracy effects in the {A}l{G}a{N}/{G}a{N} two-dimensional electron gas channel},
journal = {Semiconductor Science and Technology}
}

@article{lu2025review,
  title={A review of {G}a{N} {RF} devices and power amplifiers for 5{G} communication applications},
  author={Lu, Hao and Zhang, Meng and Yang, Ling and Hou, Bin and Martinez, Rafael Perez and Mi, Minhan and Du, Jiale and Deng, Longge and Wu, Mei and Chowdhury, Srabanti and others},
  journal={Fundamental Research},
  volume={5},
  number={1},
  pages={315--331},
  year={2025},
  publisher={Elsevier}
}

@article{meneghesso2014breakdown,
  title={Breakdown mechanisms in {A}l{G}a{N}/{G}a{N} {HEMT}s: {A}n overview},
  author={Meneghesso, Gaudenzio and Meneghini, Matteo and Zanoni, Enrico},
  journal={Japanese Journal of Applied Physics},
  volume={53},
  number={10},
  pages={100211},
  year={2014},
  publisher={IOP Publishing}
}

@article{ambacher1999two,
  title={Two-dimensional electron gases induced by spontaneous and piezoelectric polarization charges in {N}-and {G}a-face {A}l{G}a{N}/{G}a{N} heterostructures},
  author={Ambacher, Oliver and Smart, J and Shealy, JR and Weimann, NG and Chu, K and Murphy, M and Schaff, WJ and Eastman, LF and Dimitrov, R and Wittmer, L and others},
  journal={Journal of Applied Physics},
  volume={85},
  number={6},
  pages={3222--3233},
  year={1999},
  publisher={American Institute of Physics}
}

@article{barker2005bulk,
  title={Bulk {G}a{N} and {A}l{G}a{N}/ {G}a{N} heterostructure drift velocity measurements and comparison to theoretical models},
  author={Barker, JM and Ferry, DK and Koleske, DD and Shul, RJ},
  journal={Journal of Applied Physics},
  volume={97},
  number={6},
  year={2005},
  publisher={AIP Publishing}
}

@article{matulionis2006hot,
  title={Hot phonons in {G}a{N} channels for {HEMT}s},
  author={Matulionis, A},
  journal={physica status solidi (a)},
  volume={203},
  number={10},
  pages={2313--2325},
  year={2006},
  publisher={Wiley Online Library}
}

@article{tunga2022comparison,
  title={A comparison of a commercial hydrodynamics TCAD solver and Fermi kinetics transport convergence for {G}a{N} {HEMT}s},
  author={Tunga, Ashwin and Li, Kexin and White, Ethan and Miller, Nicholas C and Grupen, Matt and Albrecht, John D and Rakheja, Shaloo},
  journal={Journal of Applied Physics},
  volume={132},
  number={22},
  year={2022},
  publisher={AIP Publishing}
}

@article{grupen2019reproducing,
  title={Reproducing {GaN HEMT} kink effect by simulating field-enhanced barrier defect ionization},
  author={Grupen, Matt},
  journal={IEEE Transactions on Electron Devices},
  volume={66},
  number={9},
  pages={3777--3783},
  year={2019},
  publisher={IEEE}
}

@article{miller2022experimentally,
  title={Experimentally validated gate-lag simulations of {AlGaN/GaN} HEMTs using Fermi kinetics transport},
  author={Miller, Nicholas C and Grupen, Matt and Islam, Ahmad E and Albrecht, John D and Frey, Dave and Young, Richard and Lindquist, Miles and Green, Andrew J and Walker, Dennis E and Chabak, Kelson D},
  journal={IEEE Transactions on Electron Devices},
  volume={70},
  number={2},
  pages={435--442},
  year={2022},
  publisher={IEEE}
}

@INPROCEEDINGS{7548427,
  author={Miller, Nicholas C. and Albrecht, John D. and Grupen, Matt},
  booktitle={2016 74th Annual Device Research Conference (DRC)}, 
  title={Large-signal {RF} {G}a{N} {HEMT} simulation using {F}ermi {K}inetics {T}ransport}, 
  year={2016},
  volume={},
  number={},
  pages={1-2}}

@article{10.1063/1.3270404,
    author = {Grupen, Matt},
    title = {An alternative treatment of heat flow for charge transport in semiconductor devices},
    journal = {Journal of Applied Physics},
    volume = {106},
    number = {12},
    pages = {123702},
    year = {2009},
    month = {12}}

@article{grupen2011energy,
  title={Energy transport model with full band structure for {G}a{A}s electronic devices},
  author={Grupen, Matt},
  journal={Journal of {C}omputational {E}lectronics},
  volume={10},
  pages={271-290},
  year={2011},
  publisher={Springer}
}

@article{1637547,
  author={Palacios, T. and Suh, C.-S. and Chakraborty, A. and Keller, S. and DenBaars, S.P. and Mishra, U.K.},
  journal={IEEE Electron Device Letters}, 
  title={High-performance {E}-mode {A}l{G}a{N}/{G}a{N} {HEMT}s}, 
  year={2006},
  volume={27},
  number={6},
  pages={428-430}}

@incollection{ridley2013quantum,
    author = {Ridley FRS, B. K.},
    isbn = {9780{LO} mo77214},
    title = {Hot phonons},
    booktitle = {Quantum Processes in Semiconductors},
    publisher = {Oxford University Press},
    year = {2013},
    month = {08},
    doi = {10.1093/acprof:oso/9780{LO} mo77214.003.0014},
}

@article{Ridley_1996,
year = {1996},
month = {sep},
volume = {8},
number = {37},
pages = {L511},
author = {B K Ridley},
title = {The {LO} phonon lifetime in {G}a{N}},
journal = {Journal of Physics: Condensed Matter},
doi = {10.1088/0953-8984/8/37/001},
url = {https://dx.doi.org/10.1088/0953-8984/8/37/001},
}

@article{PhysRevB.68.035338,
  title = {Hot-phonon temperature and lifetime in a biased $\mathrm{{A}l_{x}{G}a_{1-x}{N}/{G}a{N}}$ channel estimated from noise analysis},
  author = {Matulionis, A. and Liberis, J. and Matulionien\ifmmode \dot{e}\else \.{e}\fi{}, I. and Ramonas, M. and Eastman, L. F. and Shealy, J. R. and Tilak, V. and Vertiatchikh, A.},
  journal = {Phys. Rev. B},
  volume = {68},
  issue = {3},
  pages = {035338},
  numpages = {7},
  year = {2003},
  month = {Jul},
  publisher = {American Physical Society},
  doi = {10.1103/PhysRevB.68.035338},
  url = {https://link.aps.org/doi/10.1103/PhysRevB.68.035338}
}

@article{Dyson_2009a,
doi = {10.1088/0953-8984/21/17/174204},
url = {https://doi.org/10.1088/0953-8984/21/17/174204},
year = {2009},
month = {apr},
publisher = {},
volume = {21},
number = {17},
pages = {174204},
author = {Dyson, A},
title = {Phonon–plasmon coupled modes in {G}a{N}},
journal = {Journal of Physics: Condensed Matter}
}

@ARTICLE{Marino2010,
  author={Marino, Fabio Alessio and Faralli, Nicolas and Palacios, TomÁs and Ferry, David K. and Goodnick, Stephen M. and Saraniti, Marco},
  journal={IEEE Transactions on Electron Devices}, 
  title={Effects of {T}hreading {D}islocations on {A}l{G}a{N}/{G}a{N} {H}igh-{E}lectron {M}obility {T}ransistors}, 
  year={2010},
  volume={57},
  number={1},
  pages={353-360},
  doi={10.1109/TED.2009.2035024}}

@ARTICLE{fang2012effect,
  author={Fang, Tian and Wang, Ronghua and Xing, Huili and Rajan, Siddharth and Jena, Debdeep},
  journal={IEEE Electron Device Letters}, 
  title={Effect of {O}ptical {P}honon {S}cattering on the {P}erformance of {G}a{N} {T}ransistors}, 
  year={2012},
  volume={33},
  number={5},
  pages={709-711},
  doi={10.1109/LED.2012.2187169}}

@article{matulionis2009plasmon,
    author = {Matulionis, A. and Liberis, J. and Matulionienė, I. and Ramonas, M. and Šermukšnis, E. and Leach, J. H. and Wu, M. and Ni, X. and Li, X. and Morkoç, H.},
    title = {Plasmon-enhanced heat dissipation in {G}a{N}-based two-dimensional channels},
    journal = {{A}pplied {P}hysics {L}etters},
    volume = {95},
    number = {19},
    pages = {192102},
    year = {2009},
    month = {11},
    issn = {0003-6951},
    doi = {10.1063/1.3261748},
}

@article{Ridley_Dyson_2009, title={{T}he {L}ifetime of {P}olar-{O}ptical {M}odes in {S}emiconductors}, volume={1221}, DOI={10.1557/PROC-1221-CC06-05}, journal={MRS Proceedings}, author={Ridley, Brian K and Dyson, Angela}, year={2009}, pages={1221-CC06-05}}

@article{dyson2011lifetime,
    author = {Dyson, A. and Ridley, B. K.},
    title = {The lifetime of optical phonons in a single heterostructure},
    journal = {Journal of Applied Physics},
    volume = {109},
    number = {5},
    pages = {054509},
    year = {2011},
    month = {03},
    issn = {0021-8979},
    doi = {10.1063/1.3553439},
}

@article{dyson2008phonon,
    author = {Dyson, A. and Ridley, B. K.},
    title = {Phonon-plasmon coupled-mode lifetime in semiconductors},
    journal = {Journal of Applied Physics},
    volume = {103},
    number = {11},
    pages = {114507},
    year = {2008},
    month = {06},
    issn = {0021-8979},
    doi = {10.1063/1.2937918},
}

@ARTICLE{matulionis2009ultrafast,
  author={Matulionis, Arvydas and Liberis, Juozapas and Matulioniene, Ilona and Ramonas, Mindaugas and Sermuksnis, Emilis},
  journal={Proceedings of the IEEE}, 
  title={Ultrafast {R}emoval of {LO}-{M}ode {H}eat {F}rom a {G}a{N}-{B}ased {T}wo-{D}imensional {C}hannel}, 
  year={2010},
  volume={98},
  number={7},
  pages={1118-1126},
  doi={10.1109/JPROC.2009.2029877}}

@article{liberis2014hot,
  title={Hot-phonon lifetime in $\mathrm{{A}l_{0.23}{G}a_{0.77}{N}/{G}a{N}}$ channels},
  author={Liberis, J and Ramonas, M and {\v{S}}ermuk{\v{s}}nis, E and Sakalas, P and Szabo, N and Schuster, M and Wachowiak, A and Matulionis, A},
  journal={Semiconductor Science and Technology},
  volume={29},
  number={4},
  pages={045018},
  year={2014},
  publisher={IOP Publishing}
}

@inproceedings{matulionis2009hot,
  title={Hot phonons in {I}n{A}l{N}/{A}l{N}/{G}a{N} heterostructure 2{DEG} channels},
  author={Matulionis, Arvydas and Morko{\c{c}}, Hadis},
  booktitle={Gallium Nitride Materials and Devices IV},
  volume={7216},
  pages={65--78},
  year={2009},
  organization={SPIE}
}

@article{srivastava2008origin,
  title = {Origin of the hot phonon effect in group-{III} nitrides},
  author = {Srivastava, G. P.},
  journal = {Phys. Rev. B},
  volume = {77},
  issue = {15},
  pages = {155205},
  numpages = {6},
  year = {2008},
  month = {Apr},
  publisher = {American Physical Society},
  doi = {10.1103/PhysRevB.77.155205},
  url = {https://link.aps.org/doi/10.1103/PhysRevB.77.155205}
}

@article{barman2004long,
  title = {Long-wavelength nonequilibrium optical phonon dynamics in cubic and hexagonal semiconductors},
  author = {Barman, Saswati and Srivastava, G. P.},
  journal = {Phys. Rev. B},
  volume = {69},
  issue = {23},
  pages = {235208},
  numpages = {16},
  year = {2004},
  month = {Jun},
  publisher = {American Physical Society},
  doi = {10.1103/PhysRevB.69.235208},
  url = {https://link.aps.org/doi/10.1103/PhysRevB.69.235208}
}

@article{tsen1998time,
  title={{T}ime-resolved Raman studies of the decay of the longitudinal optical phonons in wurtzite {G}a{N}},
  author={Tsen, Kong-Thon and Ferry, DK and Botchkarev, A and Sverdlov, B and Salvador, A and Morkoc, H},
  journal={Applied Physics Letters},
  volume={72},
  number={17},
  pages={2132--2134},
  year={1998},
  publisher={American Institute of Physics}
}

@article{tsen2006subpicosecond,
  title={Subpicosecond time-resolved Raman studies of {LO} phonons in {G}a{N}: Dependence on photoexcited carrier density},
  author={Tsen, Kong-Thon and Kiang, Juliann G and Ferry, DK and Morko{\c{c}}, Hadis},
  journal={Applied Physics Letters},
  volume={89},
  number={11},
  year={2006},
  publisher={AIP Publishing}
}

@ARTICLE{6022750,
  author={Sridharan, Sriraaman and Christensen, Adam and Venkatachalam, Anusha and Graham, Samuel and Yoder, P. D.},
  journal={IEEE Electron Device Letters}, 
  title={{T}emperature- and {D}oping-Dependent {A}nisotropic {S}tationary {E}lectron {V}elocity in {W}urtzite {G}a{N}}, 
  year={2011},
  volume={32},
  number={11},
  pages={1522-1524},
  doi={10.1109/LED.2011.2164611}}

@ARTICLE{7505603,
  author={Grupen, Matt},
  journal={IEEE Transactions on Electron Devices}, 
  title={{G}a{N} High Electron Mobility Transistor Simulations With Full Wave and Hot Electron Effects}, 
  year={2016},
  volume={63},
  number={8},
  pages={3096-3102},
  doi={10.1109/TED.2016.2581591}}

@article{khurgin2007hot,
  title={Hot phonon effect on electron velocity saturation in {G}a{N}: {A} second look},
  author={Khurgin, Jacob and Ding, Yujie J and Jena, Debdeep},
  journal = {Applied Physics Letters},
  volume={91},
  number={25},
  year={2007},
  publisher={AIP Publishing}
}

@article{ridley2004,
    author = {Ridley, B. K. and Schaff, W. J. and Eastman, L. F.},
    title = {Hot-phonon-induced velocity saturation in {G}a{N}},
    journal = {Journal of Applied Physics},
    volume = {96},
    number = {3},
    pages = {1499-1502},
    year = {2004},
    month = {08},
    doi = {10.1063/1.1762999}}

\end{document}